\def\Uc{{ U_{{\rm CPQMC}} }}
\def\Ud{{ U_{{\rm DQMC}} }}
\def\Ut{{ U_{{\rm TWF}} }}

\documentstyle[twocolumn,aps,psfig]{revtex}
\begin{document}

\title{Constrained-path quantum Monte Carlo simulations
of the zero-temperature, disordered two-dimensional Hubbard model}
\author{M. Enjalran$^{1,4,}$\cite{newadd},
F. H\'{e}bert$^2$, G. G. Batrouni$^2$, 
R. T. Scalettar$^1$, and Shiwei Zhang$^3$}

\address{
$^1$ Physics Department,
University of California,
Davis, California 95616}
\address{
$^2$ Institut Non-Lin\'{e}aire de Nice,
Universit\'{e} de Nice-Sophia Antipolis,
1361, route des Lucioles,
06560 Valbonne, France}
\address{
$^3$ Department of Physics,
College of William and Mary, Williamsburg, Virginia 23187.}
\address{
$^4$ Materials Research Institute, Lawrence Livermore National
Laboratory, Livermore, California 94550.
}

\date{\today}

\maketitle

\begin{abstract}
We study the effects of disorder on long-range antiferromagnetic
correlations in the half-filled, two dimensional, 
repulsive Hubbard model at $T=0$.  A mean field approach is first employed
to gain a qualitative picture of the physics and to guide our
choice for a trial wave function in a constrained path quantum 
Monte Carlo (CPQMC) method that allows for a more accurate
treatment of correlations. 
Within the mean field calculation, 
we observe both Anderson and Mott insulating 
antiferromagnetic phases.  There are transitions to a
paramagnet only for relatively weak coupling,
$U < 2t$ in the case of bond disorder, and $U < 4t$ in the case of 
on-site disorder.  
Using ground state CPQMC we demonstrate that this mean field
approach significantly overestimates magnetic order. 
For $U=4t$, we find a critical bond disorder of  
$V_{c} \approx (1.6 \pm 0.4)t$ even though
within mean field theory no paramagnetic phase is found for this
value of the interaction.
In the site disordered case, 
we find a critical disorder of
$V_{c} \approx (5.0 \pm 0.5)t$ at $U=4t$. 
\end{abstract}

\section{Introduction}

The Hubbard Hamiltonian encapsulates many of the 
most interesting qualitative many-body effects in correlated fermion
systems, notably the possibility of ordering of electron
spins and the appearance of insulating states
in systems with partially filled electronic bands.
While the original model is translationally invariant,
the introduction of disorder can alter the magnetic and
charge correlations in fundamental ways.\cite{REVIEWS,belitz}
Experimental systems whose qualitative
physics appears to involve the interplay of interactions and randomness,
possibly modeled by the Hubbard Hamiltonian,
include doped semiconductors\cite{belitz,sachdev},
thin superconducting films\cite{EXPERIMENTS,theories}, 
and silicon metal-oxide-semiconductor
field-effect transistors.~\cite{kravch}

The determinant quantum Monte Carlo (DQMC) method
has been a useful tool to simulate the Hubbard Hamiltonian,\cite{DET1,DET2}
but its application has been limited by the impossibility of
reaching low temperatures in many cases of interest, a problem which
arises when the ``Boltzmann weight'' for the fermion 
system becomes negative.\cite{DET2,SIGN}
An approach for dealing with this ``sign problem'' 
in the ground state
is the constrained path quantum Monte Carlo (CPQMC) method.\cite{CPQMC1}
In CPQMC, the sign problem is treated by imposing,
in the space of Slater determinants,  a boundary condition based on an
input trial wave function. 

The CPQMC approach has not previously been used in models with quenched
randomness.  In this paper, we apply CPQMC to
the disordered, two-dimensional (2D), ``Anderson-Hubbard'' Hamiltonian,
\begin{eqnarray}
H&=&-\sum_{\langle {\bf i},{\bf j} \rangle \sigma} 
t_{{\bf ij}}
(c_{{\bf i}\sigma}^{\dagger}c_{{\bf j}\sigma} + 
c_{{\bf j}\sigma}^{\dagger}c_{{\bf i}\sigma}) \nonumber \\
 & & + U \sum_{{\bf i}} (n_{{\bf i}\uparrow}-\frac12)
(n_{{\bf i}\downarrow}-\frac12)\nonumber\\
  & & + \sum_{{\bf i}} (\epsilon_{{\bf i}}-\mu) 
(n_{{\bf i}\uparrow}+n_{{\bf i}\downarrow}).
\label {eq:eq1}
\end{eqnarray}
Here $c_{{\bf i}\sigma}$ ($c_{{\bf i}\sigma}^{\dagger}$) are operators that
destroy (create) electrons of spin $\sigma$ on site ${\bf i}$ of 
a 2D square lattice of size $L^{2}=N.$ 
$U$ is the on-site repulsion, $\mu$ and $\epsilon_{{\bf i}}$
are the chemical potential and random site energies, respectively,
and $t_{{\bf ij}}$ is the (random) hopping energy.
Random on-site energies are chosen uniformly on $[-V_s/2,+V_s/2]$,
and $t_{{\bf ij}}$ are chosen uniformly on
$[t-V_t/2,t+V_t/2]$, 
where $V_s$ and $V_t$ are parameters that set the disorder strength.
We will choose $t=1$ to set the scale of energy,
and focus our attention on the case when the
lattice is half-filled $\langle n \rangle=1.$ 

In the absence of disorder, the half-filled Hubbard model has 
antiferromagnetic (AF) long-range order at all values of the ratio $U/t.$  
For large $U/t,$ each site of the lattice is singly occupied,
and well defined moments exist. 
AF order arises as a result of a second-order lowering 
of energy when neighboring electron spins are antiparallel.
In this strong-coupling regime, the density of states ${\cal N}(\omega)$ 
consists of
upper and lower Hubbard bands, separated by $U.$  The compressibility
$\kappa = N\partial \langle n \rangle / \partial \mu$ vanishes at
half-filling, reflecting the presence of a Mott-Hubbard gap.

At weak coupling, AF order is produced by nesting of the Fermi
surface, that is, $\epsilon({\bf k+Q}) = -\epsilon({\bf k})$,
at ${\bf Q}=(\pi,\pi)$, which results in 
a divergence of the noninteracting magnetic susceptibility.
Here, $\epsilon({\bf k}) = -2t ({\rm cos}k_x + {\rm cos}k_y)$
is the free-particle dispersion relation in the clean limit.
The density of states exhibits a 
Slater gap at half-filling, arising from this AF order, 
and again $\kappa$ vanishes.

Previous DQMC simulations have confirmed this picture of the
physics of the clean Hubbard model at half-filling,
and made these statements more quantitative.\cite{DET2,HIRSCH2}
An analysis of the effect of bond disorder, which we shall
review below, has also been performed.\cite{BDIS} However, 
for the case of site disorder, DQMC simulations have not proven possible.
In Sec.~\ref{sec:mft}, we review the mean
field treatment of the problem and consider the effects of disorder in
this limit.
The CPQMC algorithm is outlined in Sec.~\ref{sec:cp-algorithm}.  
The effects of disorder on the magnetic
correlations are presented in Sec.~\ref{sec:magn-corr}, and we close with 
a brief summary of our results in
Sec.~\ref{sec:concl}.
The Appendix presents some detailed tests of CPQMC on 
different model systems (both clean and disordered), with a particular
focus on the effect of different choices of the trial wave functions.

\section{Mean Field Approximation}
\label{sec:mft}

Mean field (MF) theory provides a useful starting point for the analysis of 
the phase diagram of the Hubbard model and, as we shall see, also
provides us with candidate trial wave functions
to use in the CPQMC simulations.  
In this approach, the interaction term is decoupled so that the 
electrons on site ${\bf i}$ of one spin species ``see'' only 
the {\it average} of the density of the other spin species.  
For zero disorder, the MF phase diagram of the 2D
Hubbard model, as a function of filling
and $U/t$, has been given by Hirsch.~\cite{HIRSCH}
In the presence of randomness, it is important to consider an unrestricted
Hartree-Fock ansatz (UHF) that allows for general site-dependent
occupations of each of the two spin species.~\cite{foot3}
Systems with electron-phonon interactions with quenched lattice
distortions have been studied in this approximation,\cite{BISHOP98}
as have the 3D Anderson-Hubbard model,\cite{logan}
and the propensity for spontaneous 
phase separation, stripe formation, and other inhomogeneous
charge distributions in the clean Hubbard and related 
models.~\cite{BISHOP98,EMERY,ZAANEN,BISHOP}
One of the purposes of this work is to see how such UHF results 
compare to those using CPQMC.

We will study the disordered 2D model in the UHF limit, 
treating bond and site
disorder separately. In order to capture the ground state 
of the model, our calculations are performed at 
$\beta = 1/T = 100,$ where $k_{B}=1.$  
It is useful to define several order parameters for the
different possible phases.
The magnitude of the $z$ component of the local moment,
\begin{equation}
M_{l} = \frac{1}{N} \sum_{\bf i} \langle |m_{\bf i}| \rangle
 = \frac{1}{N} \sum_{\bf i} \langle |n_{\bf i \uparrow}
 -n_{\bf i \downarrow}| \rangle,
\label{eq:eq5} 
\end{equation}
measures the tendency for sites to have different numbers of
up and down spin electrons.
The staggered magnetization, 
\begin{equation}
M_{s} = \frac{1}{2N} \sum_{\bf i} (-1)^{\bf i}
\langle m_{\bf i} \rangle,
\label{eq:eq6} 
\end{equation}
determines the degree of long range antiferromagnetic correlation
of these moments.
Here the notation $\langle \ldots \rangle$
represents an averaging over disorder.

It is also useful to look at charge correlations.
Two different types of metal-insulator transitions (MIT)
can occur in the Anderson-Hubbard model.  
In the Anderson MIT, the vanishing of the conductivity is driven 
primarily by the localizing effect of disorder.  A useful observable is
the inverse participation ratio, $R^{-1},$ 
\begin{equation}
 R^{-1} = \frac{1}{N} \sum_{\bf{k}, \sigma, {\bf i}}
|\psi_{\bf{k}\sigma, {\bf i}}|^{4},
\label{eq:eq7}  
\end{equation}
where the eigenstates of the Hamiltonian read 
$|\psi_{{\bf k}\sigma} \rangle = \sum_{\bf i} \psi_{{\bf k}\sigma, {\bf i}} 
|{\bf i}\sigma \rangle$.
For delocalized states, we have $\psi_{\bf{k}\sigma,{\bf i}}\approx
1/\sqrt{N}$ and  
$\lim_{N \rightarrow \infty}R^{-1} \rightarrow 0$ in the thermodynamic
limit. Meanwhile for localized states, the fermions
spread over a few sites and hence $R^{-1}$ goes 
to a finite value in the thermodynamic limit, signaling localized electrons.

In the Mott MIT, the particles are localized primarily 
by their interactions.  The insulating state
is marked by the presence of a gap in the charge spectrum at the Fermi 
surface and an associated vanishing of the charge compressibility,  
\begin{equation}
\kappa = \frac{\partial \langle N_{\rm part} \rangle}{\partial \mu} = 
\beta \left(\langle N_{\rm part}^{2} \rangle
- \langle  N_{\rm part} \rangle^{2}\right),
\label{eq:eq8} 
\end{equation}
where $N_{\rm part}$ is the total number of particles in the system,
i.e., $N_{\rm part} = N\langle n \rangle$.

In related work in 3D,\cite{logan} $M_l$
has been used to
distinguish a ``spin-glass-like'' and a ``paramagnetic''
disordered phase that both have $M_s=0$ but have
$M_{l}$ nonzero and $M_{l}$ zero, respectively.
Interestingly, in two dimensions,
we found the very simple result that $M_l=0$ whenever $M_s=0$.
That is, in a MF treatment of the half-filled two dimensional 
model it appears
that there is no phase in which local moments are present without
ordering antiferromagnetically. However, a spin-glass phase has been
observed away from half-filling in a similar 
two dimensional model applicable to the study of
La$_{2-x}$Sr$_x$CuO$_4$.~\cite{dasgupta} 
As we shall comment further below, we are unable to address
the possible spin-glass physics with QMC, and so we merely report here
that, apparently, within MF theory (MFT) the spin-glass phase is absent 
in two dimensions.
Monte Carlo studies of classical spin glasses have shown that,
as with many phase transitions, the appearence of spin-glass order is
made less likely as the dimensionality is
reduced.\cite{newfoot2a,newfoot2b}

A similar difference between two and three dimensions\cite{logan} is manifest
in the behavior of the inverse participation ratio.
We have evaluated $R^{-1}$ but find that it approaches finite values
(indicating insulating behavior)
throughout the phase diagram of the disordered 
two dimensional Hubbard model 
at half-filling.
Since $R^{-1}$ is finite and $M_l$ mimics $M_s$ throughout our phase diagram,
we will present results only for the staggered magnetization 
and compressibility.

\begin{figure}
\center
\leavevmode
\psfig{file=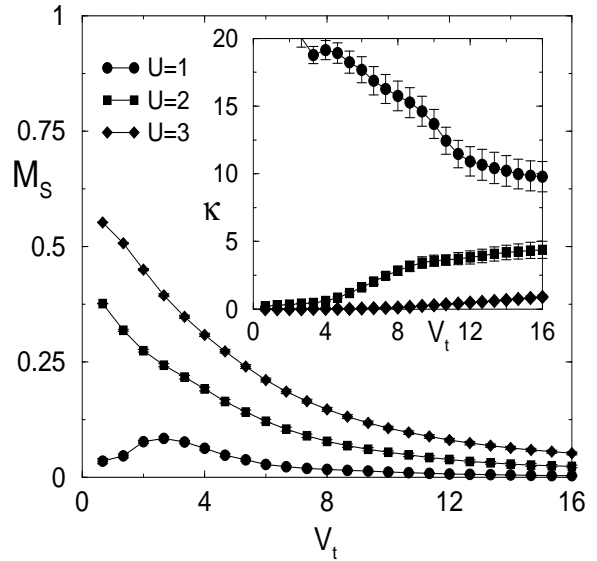,height=3.0in,width=3.0in}
\vskip03mm
\caption 
{Staggered magnetization and compressibility for the 
case of bond disorder on a $10\times10$ lattice in 
the unrestricted Hartree-Fock approximation.  AF long-range
order (LRO) is
destroyed by bond disorder for weak on-site interactions,
$U\stackrel{<}{\sim}1.0.$ The staggered magnetization, $M_{s}$,
levels off in the thermodynamic limit for $U\stackrel{>}{\sim}2.0$,
and no amount of disorder destroys AF LRO. The inset shows the
behavior of the compressibility with $V_{t}.$  A gap is present 
at $U > 2.0$ and $V_{t}<2.0,$ but it is destroyed with increasing
disorder. }
\label{fig:MsK-bond}
\end{figure}

\begin{figure}
\center
\leavevmode
\psfig{file=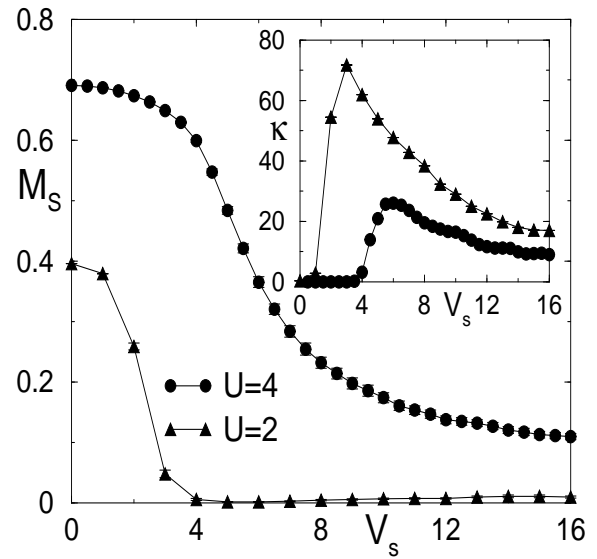,height=3.0in,width=3.0in}
\caption 
{Evolution of the staggered magnetization and compressibility
for on-site disorder on a $12\times12$ lattice. 
While there is a transition to a
paramagnetic state for $U=2$, the system remains
antiferromagnetic even at large $V_s$ for $U=4$.
The compressibility (inset) indicates that the presence of
a gap is more sensitive to site disorder than to bond disorder.}
\label{fig:SMandK}
\end{figure}

Figures.~\ref{fig:MsK-bond} and \ref{fig:SMandK} 
show these observables for sweeps
of the disorder strength at fixed values of the interaction.
In the case of bond disorder (Fig.~\ref{fig:MsK-bond}), 
the staggered magnetization
$M_s$ is nonzero at all but the smallest interactions.
The compressibility, however, has an interesting change in 
behavior as $U$ increases to $U=2t$, namely a transition
from an Anderson insulating phase with nonzero compressibility
to a Mott insulating phase with $\kappa=0$.~\cite{foot4}

For site disorder
there is a much clearer region of vanishing staggered magnetization
and hence paramagnetic behavior.  At $U=2t$, for example,
$M_s$ vanishes beyond $V_s=4t$.  
At $U=4t$, however, the physics becomes rather similar to the
bond disordered case, with AF correlation extending to
very large values of disorder, and a signature of a Mott
transition in the compressibility.
The enlargement of the paramagnetic 
phase space at the expense of the antiferromagnetically
ordered Mott and Anderson phases
is presumably a consequence of the existence,
with site disorder, of potential wells
on which pairs can form, destroying the moments. 
Similarly, we observe 
that on-site disorder is more effective at eliminating the charge 
gap than bond disorder. 

\begin{figure}
\center
\leavevmode
\psfig{file=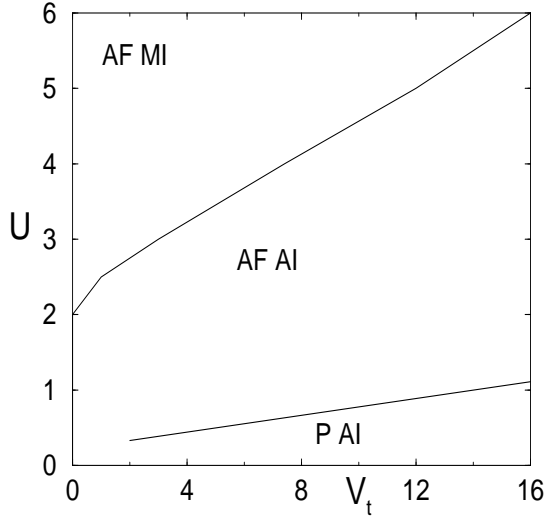,height=2.8in,width=2.8in}
\caption 
{Phase diagram of the bond-disordered Hubbard model within the
unrestricted Hartree-Fock limit. P = paramagnet, AF =
antiferromagnet, AI = Anderson insulator, MI = Mott insulator.}
\label{fig:uhfbond}
\end{figure}

\begin{figure}
\center
\leavevmode
\psfig{file=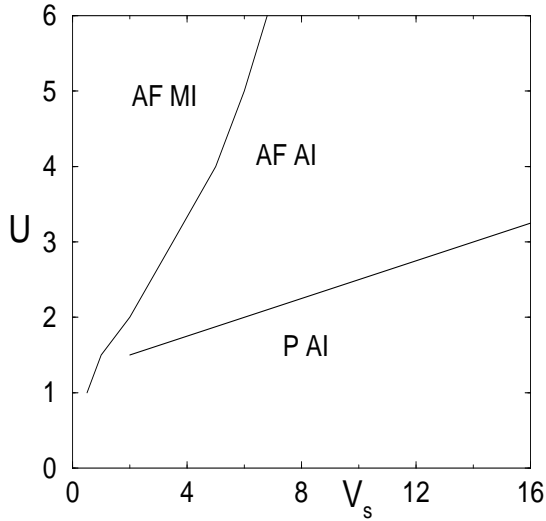,height=2.8in,width=2.8in}
\caption 
{Phase diagram of the site-disordered Hubbard model within the
unrestricted Hartree-Fock limit.  P = paramagnet, AF =
antiferromagnet, AI = Anderson insulator, MI = Mott insulator.}
\label{fig:uhfsite}
\end{figure}

We used these, and other, sweeps of disorder 
strength at fixed interaction to generate the full
UHF ground state phase diagrams of the site- and 
bond-disordered two dimensional
Hubbard models as shown in Figs.~\ref{fig:uhfbond} and
~\ref{fig:uhfsite}.
We observed both antiferromagnetically ordered
Anderson and Mott insulating phases, 
and a paramagnetic Anderson insulating region.
These three phases are described by
\begin{itemize}
\item Paramagnetic Anderson insulator (P AI): 
$M_l=0$,
$M_{s} = 0, \kappa \neq 0, R^{-1} \neq 0.$
\item Antiferromagnetic Anderson insulator (AF AI):
$M_l \neq 0$,
$M_{s} \neq 0, \kappa \neq 0, R^{-1} \neq 0.$
\item Antiferromagnetic Mott insulator (AF MI): 
$M_l \neq 0$,
$M_{s} \neq 0, \kappa = 0, R^{-1} \neq 0.$ 
\end{itemize}

In the case of bond disorder, systems of size
$6\times6$, $8\times8$, and $10\times10$ were simulated 
with averages from $40$, $50$, and $50$ disorder realizations, 
respectively.  
AF order dominates the MF phase diagram, as might be
expected since the disorder is not destroying the moments directly,
and MFT is too primitive to pick up subtle effects such as destruction of
AF order via singlet formation.
The paramagnetic region is restricted to a narrow wedge with
$V_t > 16U$.
For $U>2t=W/4$ (where $W$ is the bandwidth of the 2D tight binding model) 
we observe AF ordered phases both of
the Mott and Anderson variety with a boundary given roughly by
$U=2t+V_t/4$.
For $U<2t=W/4$ there is no Mott gap within MFT.

In the case of site disorder,
the antiferromagnetic Mott insulator exists for interaction
strengths that obey $U > V_s$.
When $U>2t$ the 
antiferromagnetic Mott insulator
is first supplanted by an 
antiferromagnetic Anderson insulator
with increasing disorder
and then ultimately by a paramagnetic Anderson insulator at much larger $V_s$.
For $U<2t$ the transition from 
antiferromagnetic Mott insulator
appears to go directly to the
paramagnetic Anderson insulator.

In summary,
a few notable results from our mean field calculation are the following: 
$(i)$ The same 
three phases, AF ordered Mott insulator, AF ordered Anderson
insulator, and paramagnetic Anderson insulator, 
are observed for bond and site disorder. 
$(ii)$ We saw no evidence for metallic ($R^{-1}=0$) or spin-glass-like
($M_l \neq 0$ with $M_s=0$) behavior.
$(iii)$ AF LRO is never destroyed by
disorder even at relatively
modest values of the interaction, e.g., $U\stackrel{>}{\sim}2t$ for bond
disorder and $U\stackrel{>}{\sim}4t$ for site disorder.   

\section{Description of the CPQMC Simulation}
\label{sec:cp-algorithm}

The ground-state CPQMC method was developed to study 
correlated lattice electrons where no special particle-hole symmetry exists
to eliminate the sign problem. 
It applies techniques that are a hybrid of
determinant Quantum Monte Carlo~\cite{DET1}
(DQMC) and
diffusion Monte Carlo~\cite{JBA1} (DMC) methods. 

Like the DQMC technique,
the method employs the Hubbard-Stratonovich (HS) transformation to 
decouple the interaction term of the Hamiltonian, 
$U\sum_{\bf{i}} n_{\bf{i}\uparrow}n_{\bf{i}\downarrow}.$ 
The result is a quadratic Hamiltonian in which the interaction 
between electrons has been replaced by the interaction of independent 
electrons with a classical fluctuating field. 
Sampling over the possible values of the HS field reproduces
the original electron-electron interaction.

This quadratic Hamiltonian can then be used in an imaginary time 
propagation of a Slater determinant, allowing projection of
the ground state from an initial trial wave function:
$|\psi_{0}\rangle = \lim_{\tau \rightarrow
\infty}\exp(-\tau H)|\psi^{(o)}\rangle$.
The similarity to the DMC method
comes from the fact that the imaginary time propagation
is represented by an ensemble of random walkers.  
(However, the random walk in 
the CPQMC method is performed in a space of Slater determinants, 
in contrast to the DMC method
where the random walk is in configuration space.)
The constrained path approximation, necessary 
for dealing with negative weights, is
similar in spirit to the fixed node~\cite{FN} approximation
commonly used to study correlated fermions in the continuum. 
It imposes a boundary condition in determinant space with a 
trial wave function,
which constrains the random walkers to half of the 
over-complete determinant space.
The details of the CPQMC algorithm have 
been discussed elsewhere~\cite{CPQMC1}, but we will provide
a brief description followed by a discussion of the necessary 
adjustments to treat disorder and the observables of interest. 

\subsection{Generation of configurations}

At any time in the CPQMC simulation, the wave function is represented
by an ensemble of random walkers.  
More specifically,
we work in a single Slater
determinant basis and represent our wave function at imaginary time
step $n$ by $|\psi^{(n)}\rangle \propto \sum_{k} |\phi^{(n)}_{k}\rangle$, 
where $|\phi^{(n)}_{k}\rangle$ is an individual walker (Slater
determinant).   The initial wave function 
$|\psi^{(o)} \rangle$ can in principle be 
any linear combination of 
Slater determinants not orthogonal to the ground state. 
For convenience, we choose it to be the unrestricted Hartree-Fock 
wave functions discussed earlier, i.e., 
$|\psi^{(o)} \rangle \equiv |\psi_{T} \rangle$.

In order to propagate the wave function forward to imaginary time $\tau$,
we discretize the propagator to a series of short time steps,
$\Delta \tau$.
This allows us to apply the Trotter
approximation 
\[
\exp(-\Delta\tau H) = \exp(-\frac{\Delta\tau K}{2})\exp(-\Delta\tau W)
\exp(-\frac{\Delta\tau K}{2}),
\] 
and isolate the potential energy $W$
from the kinetic energy $K$.
The HS transformation is then applied,  
\begin{eqnarray}
\exp(-\Delta\tau Un_{\bf{i}\uparrow}n_{\bf{i}\downarrow}) =   
\exp \left (-\frac{\Delta\tau U
(n_{\bf{i}\uparrow}+n_{\bf{i}\downarrow})}{2} \right ) 
\nonumber \\
\sum_{x_{\bf{i}} = \pm 1} p(x_{\bf{i}}) 
\exp[\gamma x_{\bf{i}} (n_{\bf{i}\uparrow}-n_{\bf{i}\downarrow})], 
\end{eqnarray}
where $\cosh(\gamma) = \exp(\Delta\tau U/2)$, $p(x_{\bf{i}})=1/2$,
and $x_{\bf{i}}$ is the HS field at site $\bf{i}$.
At each imaginary time step
the interaction part of the propagator is now a function of the HS field.

Because of the special form of the propagator,
each Slater determinant $|\phi_k^{(n)} \rangle$
in the representation of the wave function
at imaginary time step $n$
is transformed into another Slater determinant
$|\phi_k^{(n+1)} \rangle$
at imaginary time step $n+1$.
We thereby apply the incremental projection operator repeatedly to
the wave function of our system to project out the ground state.
In order to make the sampling more efficient we can
employ an importance function 
to modify the original
probability distribution, as has been discussed.\cite{CPQMC1}
As in the DQMC algorithm, the computation time in the 
CPQMC algorithm scales roughly as
$N^{3}N_{w}$ per random walk step, where $N$ is the number of spatial sites
in the lattice and 
$N_{w}$ is the number of walkers.

The CPQMC algorithm is exact up to this point.
A remaining issue is
the constraint to deal with the sign problem, which is
usually implemented with the importance
function.  We define the overlap integral
$O_{T} \equiv \langle \psi_{T}|\phi_{k} \rangle$, and demand that
individual walkers
maintain a positive $O_{T}$, i.e., that they do not cross the boundary 
$\langle \psi_{T}|\phi_{k} \rangle = 0$ in their random walk in 
Slater determinant space. 
This is applied to every walker at every time
step. The constraint is an approximation, whose quality depends on the 
quality of the trial wave function $|\psi_T\rangle$. 

\subsection{Measurements}
\label{sc:meas} 

Monte Carlo methods such as the CPQMC technique that employ trial wave functions
are well suited for calculating the ground-state energy,
and initial work on the CPQMC method\cite{CPQMC1} demonstrated
an excellent agreement of the ground state energy with
exact approaches in  cases where exact results were available.\cite{foot6}
In the Appendix,
we will demonstrate that this agreement, though a bit less
accurate, extends to our simulations.

In the body of the paper, however, we will
focus on real space magnetic order which can be identified by measuring the
correlation function,
\begin{equation}
C({\bf l})=\frac{1}{N} \sum_{{\bf j}}
\langle m_{\bf j} m_{{\bf j+l}} \rangle .
\label{eq:correl}
\end{equation}
Here $m_{\bf j}= n_{{\bf j}\uparrow}- n_{{\bf j \downarrow}}$ 
is the $z$ component of the local
spin operator, and $N$ is the total number of lattice sites.
$C(0,0)$ measures the squared local magnetic moment
$\langle m_{\bf j}^2 \rangle $.
In the clean system, $C(0,0)=0.5$ in the noninteracting limit,
and saturates at $C(0,0)=1$, as $U$ is increased.
In the clean system, 
$\langle m_{\bf j} m_{{\bf j+l}} \rangle$
is translationally invariant, that is, independent of ${\bf j}$.
For a particular disorder realization, however, this is not the case,
and translational invariance is restored only after disorder averaging.

It is useful to consider the
magnetic structure factor, the Fourier transformation of $C({\bf l})$,
\begin{equation}
S({\bf q}) = \sum_{\bf l} C({\bf l}) e^{i{\bf q}\cdot {\bf l}}.
\label{eq:struct}
\end{equation}
The structure factor will have sharp peaks at ordering vector
${\bf Q}$ when long-range magnetic order is present. 
$\beta S(0,0)$ is the uniform spin susceptibility.
At half-filling,
we always find $S({\bf q})$ to be largest at the commensurate vector
${\bf Q}=(\pi,\pi)$, even in the presence of randomness.
However, our resolution in momentum space is rather coarse and 
ordering at ${\bf Q}$ values close to $(\pi,\pi)$ 
would be difficult to see 
unless the lattice sizes were much increased. 

For finite lattice simulations,
the issue of the presence of long-range order in
the thermodynamic limit may be settled by
examining the scaling properties on lattices of different size.
Spin wave theory \cite{huse} predicts
\begin{eqnarray}
C(L/2,L/2)  & = & \frac{M^2_s}{3} + O(L^{-1}), \nonumber \\
\frac{S(\pi,\pi)}{N} & = & \frac{M^2_s}{3} + O(L^{-1}).
\label{eq:scaling}
\end{eqnarray}
Here $M_s$ is the sublattice magnetization in the thermodynamic limit,
and $(L/2,L/2)$ is the maximal separation on a square lattice of linear
size $L=\sqrt{N}$
with periodic boundary conditions. 
$C$ and $S$ provide two 
quantities to extrapolate the value of the ground state order parameter.

It is important to comment on the differences of behavior between
the order parameters based on local moments such as $M_l$ and $M_s$
[Eqs.~\ref{eq:eq5} and \ref{eq:eq6}],
which one might expect to find in comparing mean field and QMC treatments.
First, because of the relatively modest spatial lattice sizes being
simulated in the QMC scheme, over the course of a typical run the simulation
is able to explore the equivalent states that have a surplus of 
up-spin electrons
and a surplus of down-spin electrons on a given site. In the thermodynamic
limit, like a real material, the simulation could get 
stuck in one or the other,
but our lattice sizes are not large enough for that to happen. 
As a consequence,
any direct measurement we make of $M_s$ always
vanishes, and $M_s$ can only be inferred from the finite size scaling
analysis, Eq.~\ref{eq:scaling}.  
Meanwhile, MFT is able to study $M_s$ directly.
Second, MFT has a well-known tendency to over-estimate the sharpness of
the behavior of the local moment as a function of the ratio of
interaction strength to band-width.  For example, even in the clean system,
phase transitions associated with magnetic long-range order are often
associated also with abrupt formation of local moments.  However,
it is well established that within the QMC method 
the evolution of the local moment
is much less abrupt.\cite{DET2}  For both these reasons,
we do not expect to be able to observe any spin-glass (SG) 
phase transition within the QMC method, 
nor indeed do we see one.  In this work we are only able to
report that the SG transition observed in MFT at half-filling 
in the three-dimensional model apparently does not occur within the 
same MF treatment in two dimensions. Spin-glass order in interacting
Fermi systems has generated much interest recently, but most investigations
have relied on field theoretic and renormalization-group-type  
techniques.~\cite{fermiglass}

The effect of disorder on the size or
existence of the Mott gap in the CPQMC method is an interesting
question that will not be considered in detail in this 
work.  We have measured the density as a function
of chemical potential\cite{GGBANDME} and the compressibility,
and find the Mott gap is rather strongly reduced by site disorder
and considerably less affected by bond disorder,\cite{ULMKE}
but we leave a detailed analysis to a later presentation.

\section{Results for Magnetic Correlations} 
\label{sec:magn-corr}

In this section we address the primary point of interest
in the paper, the effect disorder has on the long-range 
magnetic order of the half-filled 2D Hubbard model. In particular,
we want to determine the critical disorder strength necessary to 
destroy the magnetically ordered ground state and ascertain the accuracy of
the UHF phase diagram. 
We concentrate on the case where $U=4t$, 
where the UHF shows no transition to a paramagnetic order.
This antiferromagnetic to 
paramagnetic transition can only take place in the thermodynamic
limit; hence we resort to finite size scaling techniques to calculate
the disorder strength at which AF LRO is lost. In the case of bond 
disorder, these questions have been investigated previously
with the DQMC method since there is no sign problem.~\cite{BDIS} 
We re-address these questions here in order to benchmark 
the accuracy of the CPQMC method. After the results for bond disorder
are discussed, we consider the case of random site energies and its
ability to drive the model from an antiferromagnet to a paramagnet. 
We again point out that the site disordered problem has not yet been
studied with th QMC method because of the sign problem.

\subsection{Bond disorder}  

The mechanism by
which bond disorder destroys AF LRO is 
the formation of spin-0 singlets.  When the disorder 
becomes large, the lattice contains 
strong bonds where it is favorable for nearest neighbors 
to form singlets rather than participate in AF LRO. 
Unlike the case of site disorder,
one would expect the persistence
of local correlations, $C(0,0),$ even in the paramagnetic phase. 

In our simulations of this problem, we draw random near-neighbor
hopping strengths from a 
uniform distribution about a mean value of $\langle t_{{\bf ij}}
\rangle = t = 1,$ e.g., $[1-V_{t}/2,1+V_{t}/2].$
We considered simulations with a renormalized interaction,
$\Uc \ne U$, chosen, as described in the Appendix,
so the CPQMC and DQMC results match in the clean limit. For $4\times4$, 
$6\times6$, and $8\times8$ lattices the values of 
$\Uc$ were set
to $1.25$, $1.75$, and $1.85$, respectively. Disorder averaging was
performed on $10$ realizations for each disorder strength on 
each lattice.  We used as $|\psi_T\rangle$
an UHF trial wave function (TWF) obtained 
with $\Ut = 2$, but as shown in the Appendix, the results
are not sensitive to this choice.\cite{foot8}

\begin{figure}
\center
\leavevmode
\psfig{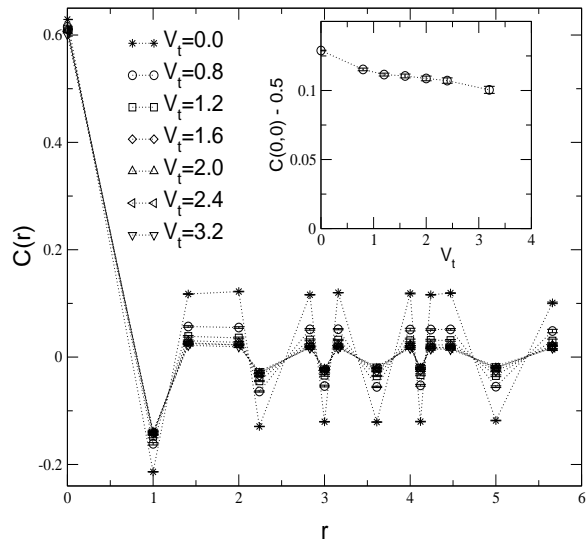}
\vskip+02mm
\caption
{The real space spin-spin correlations on a bond-disordered $8 \times 8$
lattice.  The inset shows
the squared local moment scaled by the noninteracting value as a function of 
bond disorder strength.}
\label{fig:CrTU2}
\end{figure}

The real space spin correlations are shown in Fig.~\ref{fig:CrTU2} 
on a fixed $8\times8$ lattice 
size.  Disorder substantially
decreases the antiferromagnetic order, with only
a relatively minor suppression of the squared local moment (inset).

\begin{figure}
\center
\leavevmode
\psfig{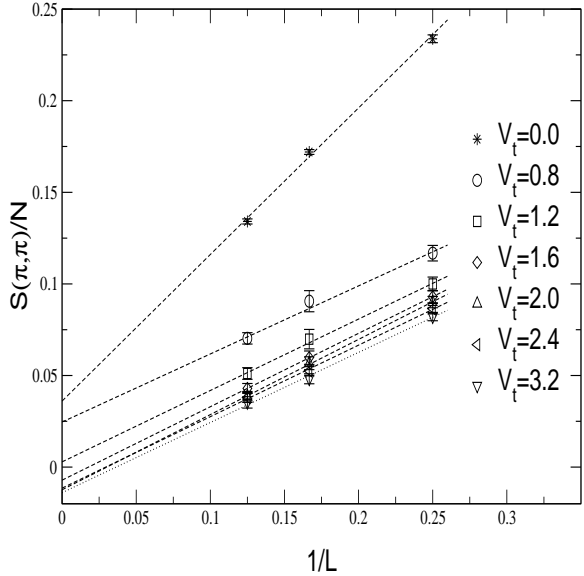}
\vskip+02mm
\caption
{Finite size scaling for the structure factor as a function of
bond disorder. Simulations were performed with a renormalized
interaction and an UHF trial state with $\Ut = 2$.
Extrapolations to the thermodynamic limit give an intercept that is equal to 
$M_{s}^{2}/3$. A critical disorder strength of $V_{t} \approx 1.6 \pm
0.4$ is found, which agrees  with the DQMC result.
The dashed lines are linear least squares fits to the data.}
\label{fig:SqNTU2}
\end{figure}

Summing up these real space
spin correlations yields the magnetic structure factor.
Working on a range of lattice sizes, and employing 
the scaling relationship 
for $S(\pi,\pi)$, Eq.~\ref{eq:scaling},
yields the staggered magnetization as a function of disorder 
from the intercepts of our scaling plot, Fig.~\ref{fig:SqNTU2}.

\begin{figure}
\center
\leavevmode
\psfig{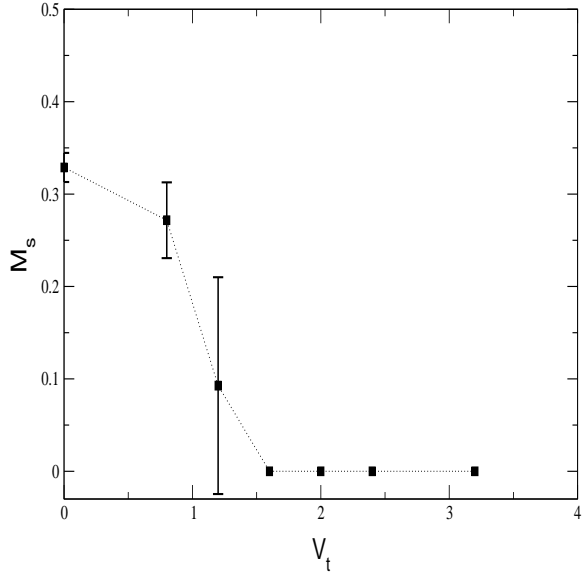}
\vskip+03mm
\caption 
{Staggered magnetization as a function of
bond disorder. Values were calculated from the intercepts of 
Fig.~\ref{fig:SqNTU2}. 
$M_{s}$ vanishes at the critical disorder strength $V_{t} \approx 1.6 \pm
0.4$.}
\label{fig:MsTU2}
\end{figure}

It is worth making two further comments conncerning Fig.~6.
First, in papers establishing long-range antiferromagnetic
order in the clean, half-filled, two-dimensional Hubbard model,
extrapolations of both the longest-range 
spin-spin correlation function $C(L/2,L/2)$ and the structure factor
$S(\pi,\pi)$ were done.\cite{DET2}
The error bars on $C$ were only
slightly larger than those from $S$, so they provided essentially
equivalent (and consistent) information.  In the disordered system, however,
we have found that $C$ is not as useful to examine as the structure
factor.  We believe that the reason is that on a lattice of $N$ sites, 
$C$ is obtained through an average over only the $O(N)$ pairs of sites
separated by $(L/2,L/2)$ while $S$ involves all $O(N^2)$ separations.
The structure factor is thus much less sensitive to individual
disorder realizations.  
Second, the linear extrapolation to negative order parameter 
in the disordered phase evident in Fig.~\ref{fig:SqNTU2} has been observed
in previous simulations, e.g., of the clean periodic Anderson model
as the hybridization between localized and delocalized orbitals is
increased until the antiferromagnetism gives way to Kondo singlet
formation.\cite{PAM}  The point is that Eq.~\ref{eq:scaling},
which yields a linear extrapolation in $1/L$, is valid only in the ordered
phase.  In the disordered phase $S/N \propto b/L^2$ so one should
actually fit the data to a quadratic expression that goes through the origin.

The results for $M_s$ are exhibited in Fig.~\ref{fig:MsTU2},
which gives a critical disorder of $V_{t} = 1.6 \pm 0.4$.
The UHF calculation predicts no transition to a paramagnetic phase for
this value of $U$,
but the CPQMC calculation agrees very well with the previous 
DQMC results.~\cite{BDIS}  
We did not observe an enhancement of $M_{s}$ at weak disorder
as was observed in our UHF data and in DMFT and DQMC 
calculations,\cite{ULMKE,BDIS}
but our resolution might be too small to observe this effect.

\subsection{Site disorder}

In our simulations of the site-disordered model, 
random energies were selected from a uniform distribution
$\epsilon_{\bf i} \in [-V_{s}/2,V_{s}/2]$.  Sites with 
$\epsilon_{\bf i} < 0$ favor double occupancy while sites with
$\epsilon_{\bf i} > 0$ favor the unoccupied state. 
This leads to a direct destruction of moments,
unlike the case of bond disorder.
In the presence of 
a repulsive Hubbard interaction $U$, there is therefore 
a competition between a lattice
with local moments and AF LRO, which is favored by $U$, 
and a state of doubly occupied and empty
sites favored by the disorder.  

\begin{figure}
\center
\leavevmode
\psfig{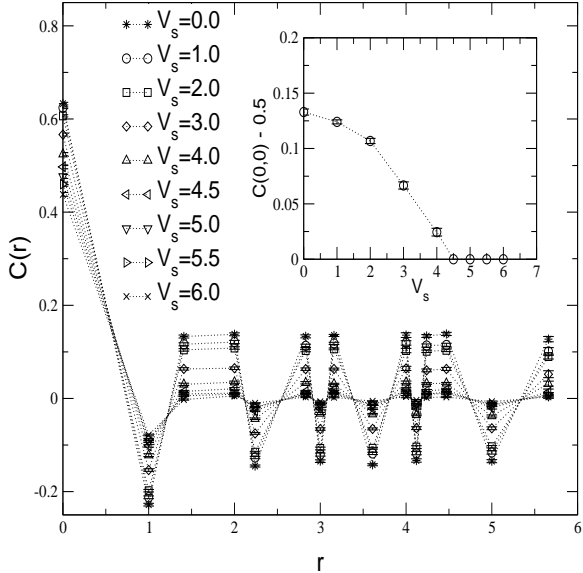}
\vskip+03mm
\caption 
{The effect of site disorder on the real space spin-spin 
correlations on an $8\times8$ lattice. Correlations for
distances greater than $1$ are uniformly reduced by disorder.
The inset shows the behavior of the scaled squared local moment
as a function of $V_{s}$. Site disorder strongly suppresses the 
local moment.}
\label{fig:Cr1}
\end{figure}

As discussed in the Appendix,
simulations with on-site disorder need to have $\Ut > V_{s}$ 
in order to capture the physics of the model and not the effect of
trial wave function. 
We used a trial wave function with $\Ut = 6$, and
the same renormalized interaction $\Uc$ used 
in the bond-disordered case.  We simulated
$10$ realizations of each disorder strength. 
The suppression of the real space spin correlations is
displayed in Fig.~\ref{fig:Cr1}.

\begin{figure}
\center
\leavevmode
\psfig{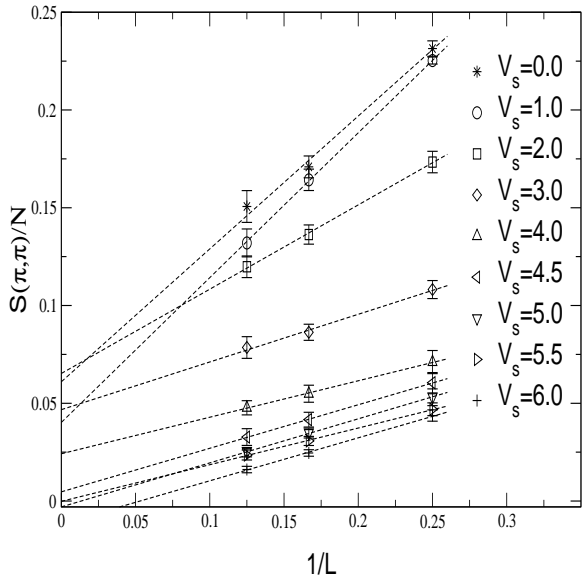}
\vskip+03mm
\caption 
{Scaling relationship for the site disordered model with 
a renormalized $\Uc$ and an UHF trial state with
$\Ut=6$. The critical disorder
strength was $V_{s} \approx 5.0 \pm 0.5$.
The dashed
lines are linear least-squares fits to the data.}
\label{fig:SqNVU6}
\end{figure}

\begin{figure}
\center
\leavevmode
\psfig{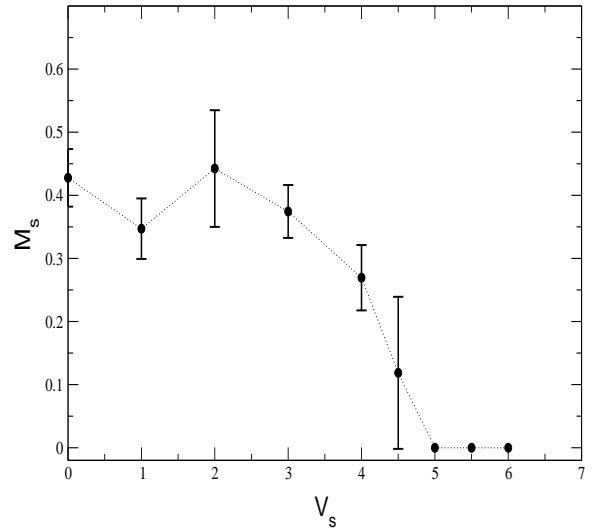}
\vskip+03mm
\caption 
{Staggered magnetization ($M_{s}$) for the site-disordered 
2D Hubbard model. Data were obtained from the intercepts of 
the scaling relationship for $S(\pi,\pi)$,
Fig.~\ref{fig:SqNVU6}. $M_{s}$ vanishes at 
$V_{s} \approx 5.0 \pm 0.5.$}
\label{fig:Ms1}
\end{figure}

We can again analyze appropriately scaled data on different lattices,
with the results shown in Fig.~\ref{fig:SqNVU6}.
The intercepts give the staggered magnetization,
which is driven to zero above a critical site
disorder strength of $V_{s} \approx 5.0 \pm 0.5$ 
(Fig.~\ref{fig:Ms1}).
The UHF calculation predicts no transition to a paramagnetic phase for
this value of $U$.  Hence, our UHF trial  wave function
has long-range antiferromagnetic correlations throughout the range of
$V_s$ in Fig.~\ref{fig:Ms1}, and the destruction of order is 
not associated with any change in the nature of $|\psi_T\rangle$.

\begin{figure}
\center
\leavevmode
\psfig{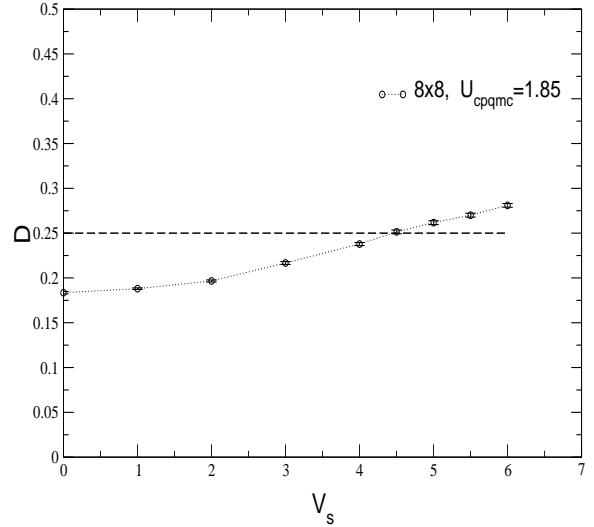}
\vskip+03mm
\caption 
{The double occupancy as a function of site disorder on 
an $8\times8$ lattice. The dashed line denotes the separation between
the repulsive (below) and attractive (above) Hubbard model at $t=0$
and any temperature. Data were obtained from simulations with 
$\Uc=1.85$ and $\Ut=6$.}
\label{fig:DV}
\end{figure}

It is interesting that the point at which AF LRO
is lost corresponds rather closely to the value of
randomness where 
the squared local moment is
reduced below its non-interacting value.
This is emphasized in Fig.~\ref{fig:DV}, which shows 
the average double occupancy 
$D = \langle n_{\uparrow} n_{\downarrow} \rangle 
= E_{I}/\Uc$ in the CPQMC method, where $E_I$ is the interaction energy
of the fermions. 
Since $C(0,0)=1-2D$, an enhancement of $D$ above the noninteracting
value is synonymous with the moment falling below the
$U=0$ value.

\section{Conclusions}
\label{sec:concl}

We have studied the ground state half-filled 2D Hubbard model 
with both bond and site
disorder in the mean field limit and by the constrained 
path quantum Monte Carlo method.  
Our most significant quantitative result was 
the first computation of the
critical disorder strength for the destruction of antiferromagnetic
long-range order by site randomness,
$(V_s)_{\rm crit}=5t$ for $U=4t$.
For this value of the interaction strength, no amount of site disorder
destroys the order in the Hartree-Fock approach, so
this emphasizes the need for better treatment of correlations
that techniques like the quantum Monte Carlo method provide.
In general, we find that
UHF calculations grossly overestimate the tendency 
for magnetic order when compared to the CPQMC calculations,
as might be expected by an approach which ignores
fluctuations.

There is less significant disagreement between UHF and CPQMC 
techniques for
the transport properties.  Unlike the 3D case, the UHF 
treatment finds that
the 2D Hubbard model is insulating for all values of
interaction and disorder at half-filling:
the inverse participation ratio was always nonzero.  This is consistent
with recent QMC calculations in two dimensions.\cite{DISHUBQMC}
A further difference between the 2D and 3D UHF results is our
conclusion that the
spin-glass phase is absent in two dimensions at half-filling.
It is interesting to note that there have been 
some indications in the QMC treatment of a metal-insulator transition
{\it off half-filling} in the 2D Hubbard model.\cite{DENT}

Our work further evaluated
and extended the range of validity of the CPQMC method by applying 
it to random systems.  We found that, as is the case for clean systems,
the CPQMC technique can provide an accurate way of treating the Hubbard model.
In particular, it gives the same critical disorder strength as
the DQMC method in the case when DQMC has no sign problem
(bond randomness), which
provides some confidence in applying it to cases such as site-disordered
problems where the DQMC method cannot give reliable results.

Much of the initial theoretical evidence for, and understanding of,
questions of charge ordering in Hubbard-like models has come from 
UHF treatments.\cite{ZAANEN}
Recent work with techniques such as the density matrix
renormalization group have emphasized that stripe formation is
a subtle and delicate effect.\cite{WHITE}
Our work indicates that there are significant
corrections to the spin correlations within UHF treatments,
and that further CPQMC calculations 
hold promise to shed some light on the behavior of disordered and
interacting electron systems.

\acknowledgements
We would like to thank Malvin Kalos for many useful discussions. 
M.~E. would also like to acknowledge the support of the Material 
Research Institute at Lawrence Livermore National Laboratory and
the Institut Non-Lin\'{e}aire de Nice-Sophia Antipolis, Valbonne, 
France for their hospitality during a visit where valuable work on
this collaboration was conducted.    
This work was supported by the Materials Research Institute of 
LLNL, by NSF-DMR-9985978, and also by NSF-DMR-9734041. S.~Z. is 
a Research Corporation Cottrell Scholar. Work at Lawrence
Livermore National Laboratory performed under the auspices of
the U.S. Department of Energy under contract No.: W-7405-ENG-48.

\vskip0.2in
\centerline{{\bf APPENDIX:  TESTS OF THE ALGORITHM}}
\vskip0.2in

In this Appendix we describe some tests of
the CPQMC algorithm with a specific focus on the effect of the
trial wave functions on the results.
Previously, the CPQMC algorithm  
has been extensively tested for
interacting systems with no disorder. Investigations have been performed on
the single-band Hubbard model in regions of parameter space where a
severe sign problem is known to exist.~\cite{CPQMC1} The method has
also been used to the study superconductivity in the 2D Hubbard model
by looking for long-range pairing correlations in the ground 
state\cite{CPQMC2} and for ferromagnetism in the periodic Anderson
model.\cite{CPQMC3}
Here disorder has been considered within a CPQMC 
calculation.

CPQMC is an exact algorithm, for all observables, 
in the absence of interactions,
even when disorder is turned on.
As a check of our code, we therefore first verified that the CPQMC code
reproduced results from exact diagonalization (ED)
at different disorder strengths.  Likewise, the DQMC algorithm
agrees perfectly with ED results.\cite{foot7}

We next looked in detail at the behavior
of the magnetic structure factor as a function of
disorder and interaction strengths, and as a function
of the trial wave function.
In the following discussion it is useful to
distinguish between the
value of the interaction strength, $\Uc$, used in the 
CPQMC algorithm, the value of the interaction strength,
$\Ut$, used in the trial wave function, and the 
physical value of $U$ in the Hamiltonian. We concentrate here on
the case where $U=4t$ and $t=1$.
In our DQMC simulations, which we used to provide benchmarks
for the CPQMC method, we of course always choose 
$\Ud=U$ since the DQMC treatment is exact.\cite{foot7}

\begin{figure}
\psfig{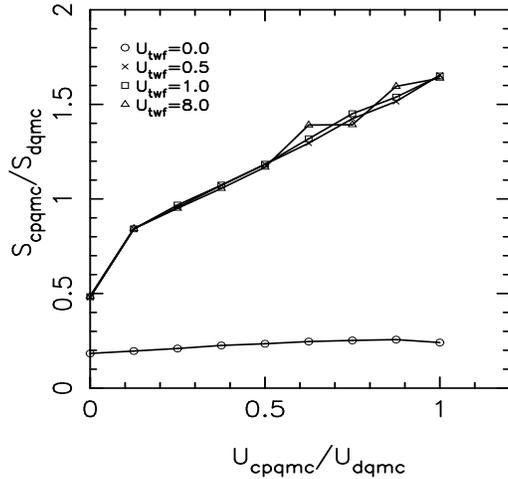}
\caption
{Scaled data for $S(\pi,\pi)_{CPQMC}/S(\pi,\pi)_{DQMC}$ as a
function of $\Uc/\Ud$ for the clean Hubbard model in the
CPQMC treatment.  When $\Uc= \Ud (=U)$, AF LRO is considerably overestimated.
We found agreement between the two methods at 
$\Uc \approx 1.25$ for this $4\times4$ lattice.
The interaction strength $\Ut$ of the UHF trial wave function 
does not affect results as long as $\Ut \neq 0$.
}
\label{fig:scale-clean}
\end{figure}

The key result of our studies
is that the CPQMC technique with UHF trial wave functions significantly overestimates
the magnetic correlations if $\Uc=U$.
This is illustrated in Fig.~\ref{fig:scale-clean},
which shows the ratio of the CPQMC to DQMC structure factors 
for the clean system on a $4\times4$ lattice as a function of $\Uc/U$.
The results are independent of the value of $\Ut$ as long as 
$\Ut \neq 0$. The structure factor is the same for the two methods
when $\Uc/U \approx 0.31$ or $\Uc \approx 1.25.$ For $6\times6$
and $8\times8$ lattices agreement is attained at $\Uc \approx 1.75$
and $\Uc \approx 1.85$, respectively, for $U=4$.   

\begin{figure}
\begin{picture}(200,600)(0,0)
\put(0,405){
	\psfig{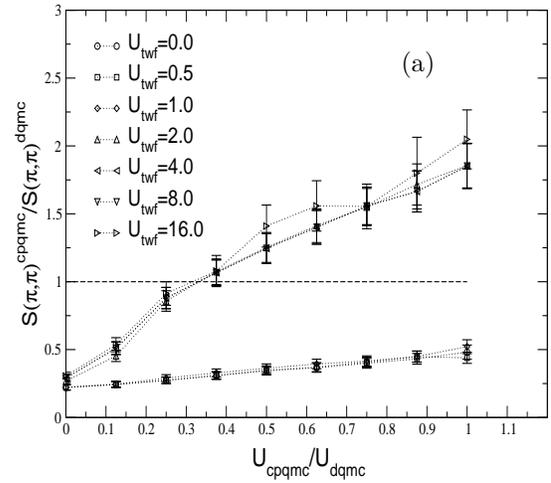}
 }
 \put(150,560){(a)}
 \put(0,205){
	\psfig{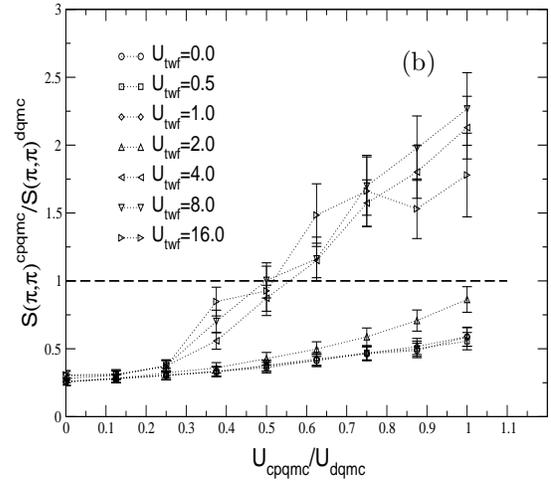}
 }
 \put(150,360){(b)}
 \put(0,5){
	\psfig{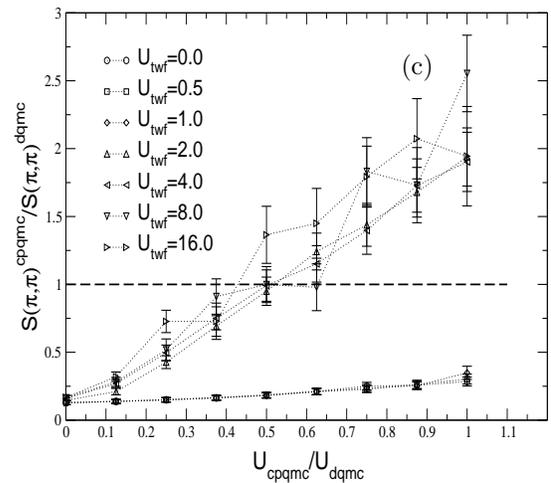}
 }
 \put(150,160){(c)}
 \end{picture}
\caption 
{Scaled data for $S(\pi,\pi)_{CPQMC}/S(\pi,\pi)_{DQMC}$ for
the site-disordered Hubbard model in the CPQMC technique: 
(a) $4\times4$, $V_{s}=2$; 
(b) $4\times4$,
$V_{s}=4$; 
(c) $6\times6$, $V_{s}=2$.}
\label{fig:scaleSppsdis}
\end{figure}

A similar effect is seen when disorder is turned on,
as illustrated in Fig.~\ref{fig:scaleSppsdis} for site disorder.
Here we show the ratio of the CPQMC and DQMC structure
factors as a function of $\Uc/U$ for different
lattice sizes.  These results are for a single
realization of disorder, which is kept fixed
as $\Uc$ and $\Ut$ are varied.  
The values of the CPQMC structure factor for different $\Ut$ fall
onto  two  curves: For all $\Ut > V_s$, the 
CPQMC method gives the same significantly overestimated structure factor.
Meanwhile, for all $\Ut < V_s$, the
CPQMC methods gives the same significantly underestimated structure factor.

These different behaviors 
are a direct manifestation of the trial wave functions.
In the unrestricted Hartree-Fock calculation,
both $\Ut$ and $V_s$ act as one-body potentials and compete with each
other: When $\Ut < V_s$, $V_s$ dominates and double
occupancy is allowed, i.e., the system is more free-electron like; when
$\Ut > V_s$, $\Ut$ dominates, double occupancy is discouraged, and
the system prefers to be in an AF state. Clearly, the 
CPQMC technique could not
adequately eliminate the biases that the two different
classes of UHF trial wave functions introduce 
through the approximate constraint to bring quantitative 
agreement between the two sets of results.
In all the work reported in the body of this paper we  chose $\Ut > V_s$
so that the trial  wave function had long-range antiferromagnetic order.
The destruction of order as randomness increased therefore must occur 
from correlation effects and not from any transition in the
trial wave function.

\begin{figure}
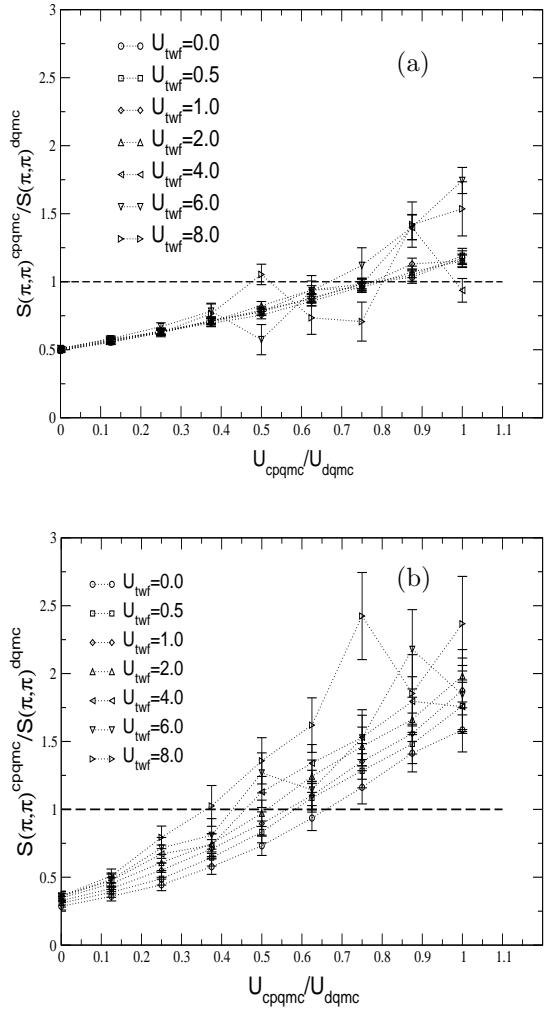

\begin{picture}(200,400)(0,0)
 \put(0,205){
	\psfig{file=fig14a.eps,height=2.5in,width=2.8in,angle=-90}
 }
 \put(150,360){(a)}
 \put(0,5){
	\psfig{file=fig14b.eps,height=2.5in,width=2.8in,angle=-90}
 }
 \put(150,165){(b)}
 \end{picture}
\caption
{Scaled data for $S(\pi,\pi)_{CPQMC}/S(\pi,\pi)_{DQMC}$ for
the bond-disordered Hubbard model in the CPQMC method: 
(a) $4\times4$, $V_{t}=2$;
(b) $6\times6$, $V_{t}=2$.}
\label{fig:scaleSppbdis}
\end{figure}

Bond disorder is studied in Fig.~\ref{fig:scaleSppbdis}.
The CPQMC structure factor is again overestimated.  
Bond disorder $V_t$, however, does
not turn into one-body potentials in the UHF method in a simple manner, 
and unlike site disorder, there are not two separate behaviors.
The result is rather independent of $\Ut$.\cite{foot8}

While we show results for single disorder realizations in 
Figs.~\ref{fig:scaleSppsdis} and \ref{fig:scaleSppbdis},
the values of the renormalized couplings used in determining, for example,
the critical disorder strength, in the main text of this paper were
obtained with disorder averaging.
The overestimation of the structure factor is a significant concern
in our studies of the destruction of long-range antiferromagnetic order,
which rely primarily on this quantity.
It exemplifies the difficulty that faces all approximate methods 
to deal with the sign problem that use a trial wave function to 
constrain the QMC sampling, namely the results can be biased
by the trial wave function, sometimes significantly.
Our approach is to fix $\Uc$ at a ``renormalized'' value
so that the structure factor from the CPQMC algorithm 
matches the DQMC result.
This sort of tuning of the interaction strength has previously been
done in comparisons of diagrammatic calculations for
the Hubbard model with DQMC results.\cite{rnU}
A crucial question, of course, is whether the
renormalized $\Uc$ is independent of lattice
size.  We found that 
$\Uc$ depends only weakly on lattice size
for $L>4$, as seen in Fig.~\ref{fig:scaleSppbdis}. 
Again, similar effects are known in the comparisons
of DQMC and diagrammatic calculations.\cite{rnU}

A further indication of the importance 
of the renormalization of the interaction lies in the behavior of
the staggered magnetization per site.
Data using a renormalized $\Uc$  always lie
below the classical upper limit of $0.5$ whereas
simulations for fixed $\Uc=U$ did not. 
At $V_{t}=0$,
our result with a renormalized $\Uc$ 
and $\Ut=2$ was $M_{s} = 0.33(2)$. 
This clean system value compares well to earlier
results for the quantum Heisenberg model obtained from a QMC
calculation, $0.30(2)$,~\cite{reger} and from 
perturbation series expansions, $0.313$.~\cite{huse} 

Another crucial question is whether the renormalized $\Uc$ depends on 
disorder strength. This question can be addressed in the case of
bond disorder where DQMC simulations of large lattices at low
$T$ can be done without encountering the sign problem,
but cannot be done for site disorder.
We found that for a given lattice size a single constant choice of $\Uc$
could be used for all $V_t$.\cite{foot8}   
We note that
the apparent variation of the renormalization on the values of $V_s$ and
$V_t$  evident in comparing Figs.~12--14 is dominantly due to the
fact the data presented there are not disorder averaged, a particularly
important issue for the smaller $4\times4$ lattices.  When such averaging is
done, as in the main body of this paper, the variation is very
significantly reduced.
This is fortunate, since 
the tuning of $\Uc$ for different disorder strength would be not
only awkward but would also call into question whether transitions
we observe as a function of disorder strength were caused by the 
tuning or by the randomness itself.

\begin{figure}
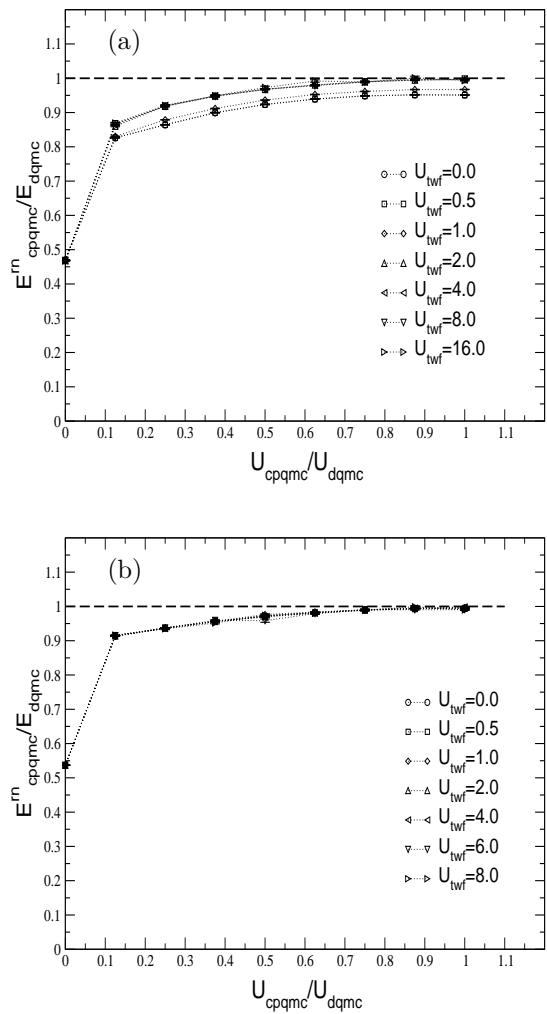

\begin{picture}(200,400)(0,0)
\put(0,205){
	\psfig{file=fig15a.eps,height=2.5in,width=2.8in,angle=-90}
}
\put(40,370){(a)}
\put(0,5){
	\psfig{file=fig15b.eps,height=2.5in,width=2.8in,angle=-90}
}
\put(40,170){(b)}
\end{picture}
\caption 
{Scaled data for the renormalized energy for the disordered 
Hubbard model in the CPQMC method: (a) $4\times4$, $V_{s}=2$; 
(b) $4\times4$, $V_{t}=2$.}
\label{fig:scaleErndis}
\end{figure}

We have focused here on the behavior of the structure factor and matching
the DQMC and CPQMC values.  Previous work has shown that the energy
and other correlations agree well.\cite{CPQMC1,CPQMC2,CPQMC3}
We have verified that the energy in DQMC and CPQMC techniques remains in
relatively good agreement in these disordered systems
if the trivial difference in interaction energy is accounted for
by defining,
\begin{equation}
E_{CPQMC}^{rn} = E_{CPQMC} + (\Ud - \Uc)(D - \frac{1}{4}).
\label{eq:rnE}
\end{equation}
Comparisons of
the energy in CPQMC and DQMC techniques behave as shown in 
Fig.~\ref{fig:scaleErndis}. For the parameters used in our 
simulations, the energies disagree by at most 5\%.

\begin{thebibliography}{99}
\bibitem{foot3}
A yet more general MF treatment
would start with a spin-rotationally invariant Heisenberg form for the
interaction, and is needed, for example to capture the fact that
$T_{\rm N\acute{e}el}=0$ for the 2D model as required by 
the Mermin-Wagner theorem.
However, since our focus is 
on ground state properties
we have followed the example of others (Ref. \cite{logan}) and
focused on the simpler decoupling,
which breaks rotational invariance.
\bibitem{newfoot2a}  
Still, it is not
completely clear why MFT should be sensitive to the dimensionality.
Perhaps the explanation lies in the special features
(perfect nesting, logarthmic divergence of the density of states)
of the clean, non-interacting band structure in two dimensions.
\bibitem{foot4}
The enhancement in 
the staggered magnetization (Fig.~\ref{fig:MsK-bond}) at small 
bond disorder and small $U=1.0$ and $1.5$ has been observed before
(Ref. \cite{BDIS})
and explained by the enhancement of the 
Heisenberg exchange with disorder.
\bibitem{foot8}
The structure factor 
of the bond-disordered model is relatively
insensitive to $\Ut$. Increasing $\Ut$
leads to larger fluctuations in the walker
population and to a tendency for the persistence of
a small staggered magnetization out to large
$V_{t}$.   Because of their lower
population fluctuations, we believe the results for small $\Ut$ are 
the more reliable ones. 
One way around this problem
would be to adjust the values of the renormalized interaction
downward in simulations with
large $\Ut$.  Indeed, in Fig.~\ref{fig:scaleSppbdis} we observe a 
more sensitive dependence of the renormalized
interaction on $\Ut$ and $V_{t}$ in larger systems.
\bibitem{foot7}
We note that the CPQMC code works at fixed particle number and $T=0$,
while the DQMC code we used was grand canonical and finite $T$.
To obtain ground state properties in DQMC we ran at different,
decreasing $T$ until the observables converged.
\end{thebibliography}
\end{document}